Energy States of Universe and New Phantom Energy

Mahgoub Salih
Department of Physics, King Saud University, Teachers College
msalih@ksu.edu.sa

Energy states of the universe is obtained when the scale factor is defined as $a = At^n$ ,and $n$ varies as $-1 \leq n \leq 1$,with the aid of the wavefunction which was constructed in , Mahgoub Salih, A Canonical Quantization formalism of curvature squared action. we found new properties of early Phantom energy, which it`s energy density increases with time while $\omega = -\frac{1}{3}$ .

After cosmological constant was invented by Einstein,1920, Friedmann solved Einstein`s equation to show that, the universe is not static. Then the acceleration universe was discovered by Hubble.

Now a day most physicists believe that most of universe is dark matter and dark energy, which caused the expansion of universe.

Various types of dark energy have been proposed, including a cosmic field associated with inflation, a different, low-energy field dubbed quintessence, and the cosmological constant, or vacuum energy of empty space. Unlike Einstein's famous fudge factor, the cosmological constant in its present incarnation doesn't delicately (and artificially) balance gravity in order to maintain a static universe, instead, it has negative pressure, that causes expansion to accelerate. The equation of state of dark energy $\omega < -\frac{1}{3}$ [1].also there are many types of dark energy with $\omega > -\frac{1}{3}$ ,such as quintessence [2], K-essence [3] and tachyonic scalar fields described by Born-Infeld (B-I) action[4].when The equation of state $\omega < -1$ ,we find the phantom energy, where phantom energy density increases with time.

In 2006 Robert J. Nemiroff suggested that a previously unknown type of energy may have dominated the early universe Called ultralight energy. Which more attractive than light.[5] .

In our previous work[6] , we constructed a wavefunction of the universe using a quadratic lagrangian form and the dynamical equation:

$$\psi(\dot{a}) = 2i\sqrt{\frac{b}{2Ab - \sin 2Ab}}\ \sin b\dot{a}\ \sin b\dot{a}$$

$$\hat{H} = \frac{\hbar^2}{4ma\dot{\gamma}}\frac{\partial^2}{\partial x^2} + \frac{a\hbar^2}{4m\dot{\alpha}}\frac{\partial^2}{\partial \acute{x}^2} + \frac{mk^2}{a}\left[\frac{1}{\dot{\gamma}} + \frac{1}{\dot{\alpha}}\right]$$

Where: $\acute{x} = \int \dot{a}^2 dt$ ; $\dot{\alpha} = -a\dot{a} - kt$ ; $\dot{\gamma} = -\dot{a} - \int \frac{\dot{a}^2+k}{a} dt$ ; $b = \frac{2|k|m}{\hbar}$ ; $C_1 = \sqrt{\frac{b}{2Ab - \sin 2Ab}}$ , the scale factor is (a) , dot, denotes derivative with respect to the cosmic time. we used the familiar scale factor relation $a = At^n$

**Energy states of universe**

Now we can extend $n$ to include negative values $-1 \le n \le 1$.

We find, $\dot{a}$ , vary as:

$$-\frac{A}{t^2} \le \dot{a} \le A \qquad 1$$

Changing the integral limits of $\dot{a}$ as in (1), to find the generalized normalization constant:

$$N = \sqrt{\frac{b}{2Ab\left(1+\frac{1}{t^2}\right) - \sin {}^{2Ab}/_{t^2} - \sin 2Ab}} \qquad 2$$

The generalized wavefunction became:

$$\psi_G(\dot{a}) = 2i\sqrt{\frac{b}{2Ab\left(1+\frac{1}{t^2}\right) - \sin {}^{2Ab}/_{t^2} - \sin 2Ab}}\ \sin b\dot{a} \qquad 3$$

One finds that $N \to C_1\ when\ t \to \infty$ , this means that this model is valid at the beginning of the universe. and construe to our model[6] at long time interval.

Using the scale factor definition, the dynamical equation can be written in the following form:

$$\hat{H} = -\frac{\hbar^2(n-1)}{4mna\dot{a}}\frac{\partial^2}{\partial \dot{a}^2} - \frac{\hbar^2(2n-1)}{4mna\dot{a}}\frac{\partial^2}{\partial \dot{a}^2} - \frac{mk^2}{a\dot{a}}\left[\frac{(n-1)}{n} + \frac{1}{a}\right] \qquad 4$$

$$\hat{H}\psi = \frac{\hbar^2(n-1)}{4mna\dot{a}}b^2\psi + \frac{\hbar^2(2n-1)}{4mna\dot{a}}b^2\psi - \frac{mk^2}{a\dot{a}}\left[\frac{(n-1)}{n} + \frac{1}{a}\right]\psi \qquad 5$$

Now we find the energy states as:

$$E(t,n) = \xi(\varepsilon - \eta) \qquad 6$$

Where: $$\xi = \frac{k^2 m}{a^2}\left(\frac{b}{2Ab\left(1+\frac{1}{t^2}\right) - \sin {}^{2Ab}/_{t^2} - \sin 2Ab}\right)\left[\frac{2n-1}{n}a - 1\right]$$

$$n \neq 0$$

$$\varepsilon = \sum_{i=1}^{\infty}(-1)^i \frac{\left({}^{2bA}/_{t^2}\right)^{2i}}{2i(2i)!}$$

$$\eta = \sum_{i=1}^{\infty} (-1)^i \frac{(2bA)^{2i}}{2i(2i)!} + \frac{1}{2}\ln|A|$$

We find that, when $t = 1$ :

$$E_{t=1} = \frac{N^2 k^2 m}{A^2}\left[1 - \frac{2n-1}{n}A\right]\ln|A|$$

Also we find $E = 0$ ,when $n = \frac{a}{2a-1}$ . this leads to $E = 0$ at radiation era. we found in previous work [6] that , cosmological constant vanishes at radiation era, when we defined it as the surface term.

the question here is , what is the relation between the cosmological constant and the curvature energy?.

The answer is simple, the cosmological constant is an energy appears due to space curvature. As we see when $t \to \infty$ , $E \propto \frac{1}{a}$ ,which is small amount, and this is the reason of the cosmological constant magnitude .

**New Phantom energy**

The density parameter Ω [6],is defined as:

$$\Omega = \frac{4b}{2Ab\left(1 + \frac{1}{t^2}\right) - \sin 2Ab - \sin {}^{2Ab}/_{t^2}} \sin^2(b\dot{a})$$

And energy density is:

$$\rho = \frac{12H^2 b}{8\pi G\left(2Ab\left(1 + \frac{1}{t^2}\right) - \sin 2Ab - \sin {}^{2Ab}/_{t^2}\right)} \sin^2(b\dot{a})$$

Where: H is Hubble parameter $H = \frac{\dot{a}}{a}$ .

When $t \to 0$ ; and $a = At^n$

Using the generalized constant of cosmological equation of state [6]

$$\omega = \frac{\dot{a} - (q+2)B + 3\frac{k}{\dot{a}}}{3\frac{k}{\dot{a}} - 3B} + \frac{\left(\frac{q}{3} + 1\right)t - H^{-1}}{t - H^{-1}} - 2$$

Where $q = -\frac{a\ddot{a}}{\dot{a}^2}$ ,the deceleration parameter, $B = \int \frac{\dot{a}^2 + k}{a} dt$ .

also, we can rewrite $\omega$ :

$$\omega = \frac{3 - 4n}{3n}$$

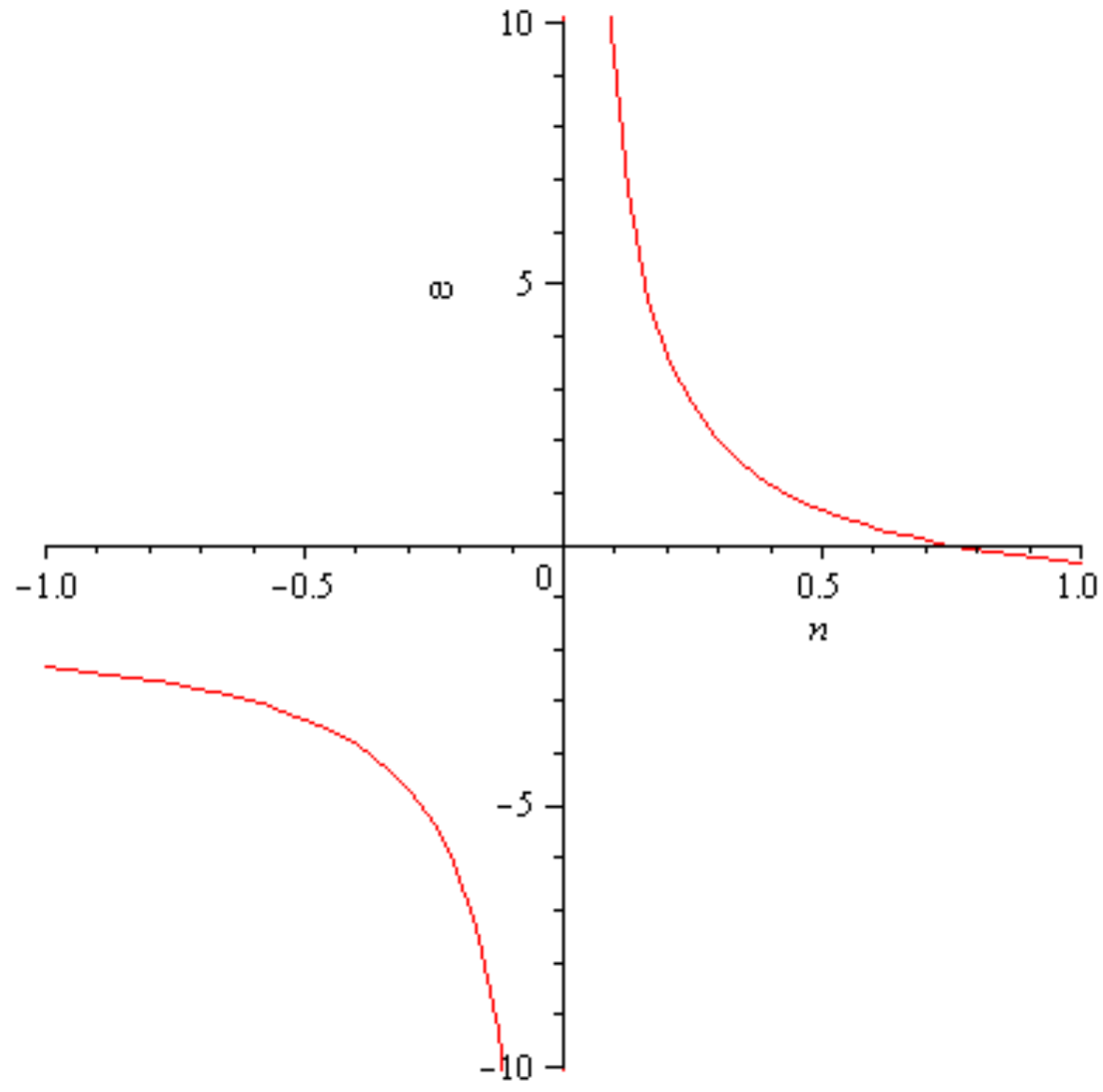

Fig(1):Shows the variations of $\omega$,with respect to $n$ .

At this early time we find the density parameter which we can identify it as Curvature density Parameter:

$$\Omega = \frac{4A^2 b^3 t^{2n}}{\frac{2Ab}{t^2} - \sin 2Ab - \sin {}^{2Ab}/{}_{t^2}} \approx 2Ab^2 t^{2(n+1)} \qquad 7$$

$$\Omega \approx \frac{4Am^2 k^2}{\hbar^2} t^{2(n+1)} \qquad 8$$

And the Curvature energy density is:

$$\rho = \frac{3n^2 A b^2 t^{2n}}{4\pi G} \qquad 9$$

When $0 < n < 1$ , energy density $\rho$ increases with time, but at this interval $\omega$ does not reach the value $-\frac{1}{3}$ ,so the universe decelerate until $n = 1$ and $\omega = -\frac{1}{3}$ . at this era ,the energy density increases with time (11).if the universe come to this era just after inflation, when $-1 \leq n < 0$ ,we find that the universe keep on expanding with constant rate and gaining energy, Phantom energy. this early time phantom energy with $\omega = -\frac{1}{3}$ , dominate the universe, A.V. Yurov [7] found the probability of universe, using Wheeler- DeWitt wave equation, after big trip, will be maximum if and only if $\omega = -\frac{1}{3}$ .

**Conclusion**:

We constructed the generalized wavefunction in interval of $-1 \leq n \leq 1$ . and obtained the energy states of the universe . we found $E = 0$ ,when $n = \frac{a}{2a-1}$ . this leads to $E = 0$ at radiation era. we calculated the Curvature density parameter $\Omega$ and the Curvature energy density $\rho$. at early universe we found a new Phantom energy, increasing with time as $\ddot{a} = 0$, with $\omega = -\frac{1}{3}$ , which dominated the universe after inflation.